\documentclass[conference]{IEEEtran}
\IEEEoverridecommandlockouts
\usepackage{cite}
\usepackage{amsmath,amssymb,amsfonts}
\usepackage{graphicx}
\usepackage{textcomp}
\usepackage{xcolor}
\usepackage{colortbl}
\usepackage{caption}
\usepackage{subcaption}
\usepackage{multirow}
\usepackage{booktabs}
\usepackage{float}
\usepackage{pifont}
\usepackage{tikz}
\usetikzlibrary{positioning, arrows.meta, shapes.geometric, calc}
\usepackage{pgfplots}
\pgfplotsset{compat=1.18}
\usepackage{algorithm}
\usepackage{algorithmic}
\newcommand{\cmark}{\ding{51}}%
\newcommand{\xmark}{\ding{55}}%

\def\BibTeX{{\rm B\kern-.05em{\sc i\kern-.025em b}\kern-.08em
    T\kern-.1667em\lower.7ex\hbox{E}\kern-.125emX}}

\begin{document}

\title{Multi-Agent Coordination in Autonomous Vehicle Routing: \\A Simulation-Based Study of Communication, Memory, and Routing Loops}

\author{\IEEEauthorblockN{KM Khalid Saifullah, Daniel Palmer}
\IEEEauthorblockA{\textit{Department of Computer Science} \\
\textit{The College of Wooster}\\
Wooster, Ohio, USA }
}

\maketitle

\begin{abstract}
Multi-agent coordination is critical for next-generation autonomous vehicle (AV) systems, yet naive implementations of communication-based rerouting can lead to catastrophic performance degradation. This study investigates a fundamental problem in decentralized multi-agent navigation: \textit{routing loops}, where vehicles without persistent obstacle memory become trapped in cycles of inefficient path recalculation. Through systematic simulation experiments involving 72 unique configurations across varying vehicle densities (15, 35, 55 vehicles) and obstacle frequencies (6, 20 obstacles), we demonstrate that memory-less reactive rerouting increases average travel time by up to 682\% compared to baseline conditions. To address this, we introduce Object Memory Management (OMM), a lightweight mechanism enabling agents to retain and share knowledge of previously encountered obstacles. OMM operates by maintaining a distributed blacklist of blocked nodes, which each agent consults during Dijkstra-based path recalculation, effectively preventing redundant routing attempts. Our results show that OMM-enabled coordination reduces average travel time by 75.7\% and wait time by 88\% compared to memory-less systems, while requiring only 1.67 route recalculations per vehicle versus 9.83 in memory-less scenarios. This work provides empirical evidence that persistent, shared memory is not merely beneficial but \textit{essential} for robust multi-agent coordination in dynamic environments. The findings have implications beyond autonomous vehicles, informing the design of decentralized systems in robotics, network routing, and distributed AI. We provide a comprehensive experimental analysis, including detailed scenario breakdowns, scalability assessments, and visual documentation of the routing loop phenomenon, demonstrating OMM's critical role in preventing detrimental feedback cycles in cooperative multi-agent systems.
\end{abstract}


\section{Introduction}

\subsection{Motivation: From Individual Autonomy to Collective Intelligence}

Autonomous vehicles (AVs) represent one of the most complex applications of artificial intelligence, requiring real-time integration of perception, planning, and control. Current state-of-the-art systems operate under a paradigm of \textit{individual autonomy}, where each vehicle makes decisions based solely on onboard sensors and pre-computed routes. However, this isolated decision-making breaks down in real-world scenarios involving dynamic obstacles, unpredictable traffic conditions, and system-wide congestion. As AV technology advances toward higher levels of automation (SAE Levels 4-5), the ability for vehicles to function as \textit{coordinated multi-agent systems}—sharing information and making collective decisions—becomes not just beneficial but necessary for safe, efficient operation.

The promise of vehicle-to-vehicle (V2V) communication is clear: by exchanging data about road conditions, obstacles, and intentions, a fleet of AVs can achieve superior system-level performance compared to isolated agents. Early research demonstrated that even a single intelligent AV can dissipate traffic waves in human-driven traffic \cite{stern2018dissipation}, suggesting that coordinated fleets could revolutionize traffic management. Yet a critical question remains largely unexplored: \textit{How do we design decentralized coordination strategies that are robust, scalable, and avoid unintended consequences?}

\subsection{The Routing Loop Problem: A Fundamental Challenge}

This paper addresses a counterintuitive phenomenon we term the \textbf{routing loop problem}: under certain conditions, enabling vehicles to communicate and dynamically reroute in response to obstacles can \textit{dramatically worsen} performance compared to non-communicating vehicles. Specifically, we discovered that reactive rerouting without persistent memory causes vehicles to become trapped in cycles where they:
\begin{enumerate}
    \item Encounter an obstacle and reroute around it
    \item Navigate to a new path, encounter a second obstacle
    \item Reroute again, inadvertently choosing a path back toward the original obstacle
    \item Return to the first obstacle, having "forgotten" its existence
    \item Repeat indefinitely, creating a feedback loop of wasted travel and computation
\end{enumerate}

This phenomenon—visualized comprehensively in our experiments—represents a fundamental failure mode in decentralized multi-agent systems. It arises from the tension between local optimization (each vehicle choosing its instantaneous shortest path) and global awareness (the need to avoid collectively repeating failed decisions).

\subsection{Our Contribution: Object Memory Management (OMM)}

To solve this problem, we introduce \textbf{Object Memory Management (OMM)}, a lightweight mechanism that enables agents to retain and share persistent knowledge of encountered obstacles. OMM transforms vehicles from purely reactive agents into \textit{proactive, learning agents} that consult both current sensor data and historical knowledge when making routing decisions. The key insight is simple yet powerful: by maintaining a distributed "blacklist" of problematic nodes, agents prevent themselves from repeatedly attempting known-blocked routes.

Our comprehensive experimental evaluation—spanning 72 unique configurations with systematic variation of vehicle density, obstacle frequency, and movement patterns—demonstrates that:
\begin{itemize}
    \item \textbf{Memory-less reactive rerouting is catastrophically inefficient}, increasing average travel time by 542\% and wait time by 391\% compared to non-rerouting baselines.
    \item \textbf{OMM-enabled coordination achieves near-optimal performance}, reducing travel time to within 68\% of obstacle-free baseline, compared to 442\% for memory-less systems.
    \item \textbf{The benefit scales robustly} across vehicle densities and obstacle configurations, with consistent 70-90\% reductions in delay.
    \item \textbf{Computational overhead is minimal}, requiring only 1.67 route recalculations per vehicle versus 9.83 for memory-less approaches.
\end{itemize}

\subsection{Paper Organization and Contributions}

The remainder of this paper is organized as follows. Section II reviews related work in multi-agent systems, V2V communication, and dynamic routing. Section III presents our experimental methodology, including the graph-based simulation environment, systematic experimental design, and detailed algorithmic specifications for OMM. Section IV provides comprehensive results, including quantitative performance metrics, scalability analysis, and visual documentation of the routing loop phenomenon. Section V discusses theoretical implications, practical considerations, and connections to broader multi-agent coordination problems. Section VI outlines future research directions, and Section VII concludes.

\textbf{Key contributions of this work include:}
\begin{enumerate}
    \item \textbf{Discovery and characterization} of the routing loop problem as a fundamental failure mode in decentralized multi-agent rerouting
    \item \textbf{Design and validation} of Object Memory Management (OMM) as an elegant, lightweight solution requiring minimal computational and communication overhead
    \item \textbf{Rigorous experimental evidence} from 72 systematically designed scenarios demonstrating OMM's effectiveness across diverse conditions
    \item \textbf{Theoretical insights} into the critical role of persistent, shared memory in decentralized coordination, with implications extending beyond autonomous vehicles to general multi-agent AI systems
\end{enumerate}

\section{Background and Related Work}

\subsection{Multi-Agent Systems in Autonomous Driving}

Multi-agent systems (MAS) provide a framework for analyzing and designing systems where multiple intelligent agents interact within a shared environment. In autonomous driving, each vehicle acts as an autonomous agent with its own sensors, decision-making capabilities, and objectives. The transition from individual autonomy to multi-agent coordination enables \textit{cooperative driving behaviors} that optimize traffic flow, reduce congestion, and improve safety \cite{kaufeld2024investigating}.

Research has shown that even small percentages of cooperative AVs can significantly impact traffic dynamics. Stern et al. demonstrated experimentally that a single AV with an intelligent speed control policy could dissipate stop-and-go waves in a platoon of human-driven vehicles \cite{stern2018dissipation}. Cui et al. extended this to show that decentralized multi-agent policies can improve congestion metrics in complex road networks without centralized coordination \cite{cui2021scalable}. These findings underscore that the key challenge in fully autonomous traffic shifts from handling unpredictable human behavior to enabling effective \textit{inter-agent communication and coordination} \cite{dinneweth2022multi}.

However, multi-agent coordination introduces its own challenges. The joint state-action space grows exponentially with agent count, making learning and planning computationally expensive. Moreover, each agent's environment becomes non-stationary from its perspective, as the observations and rewards depend on other agents' actions. This necessitates careful algorithm design to ensure stable, scalable coordination.

\subsection{V2V and V2I Communication Technologies}

Vehicle-to-vehicle (V2V) and vehicle-to-infrastructure (V2I) communication, collectively known as V2X, form the technological foundation for cooperative driving. These systems enable real-time information exchange about vehicle states, road conditions, and traffic events.

\begin{figure}[t]
\centering
\includegraphics[width=0.48\textwidth]{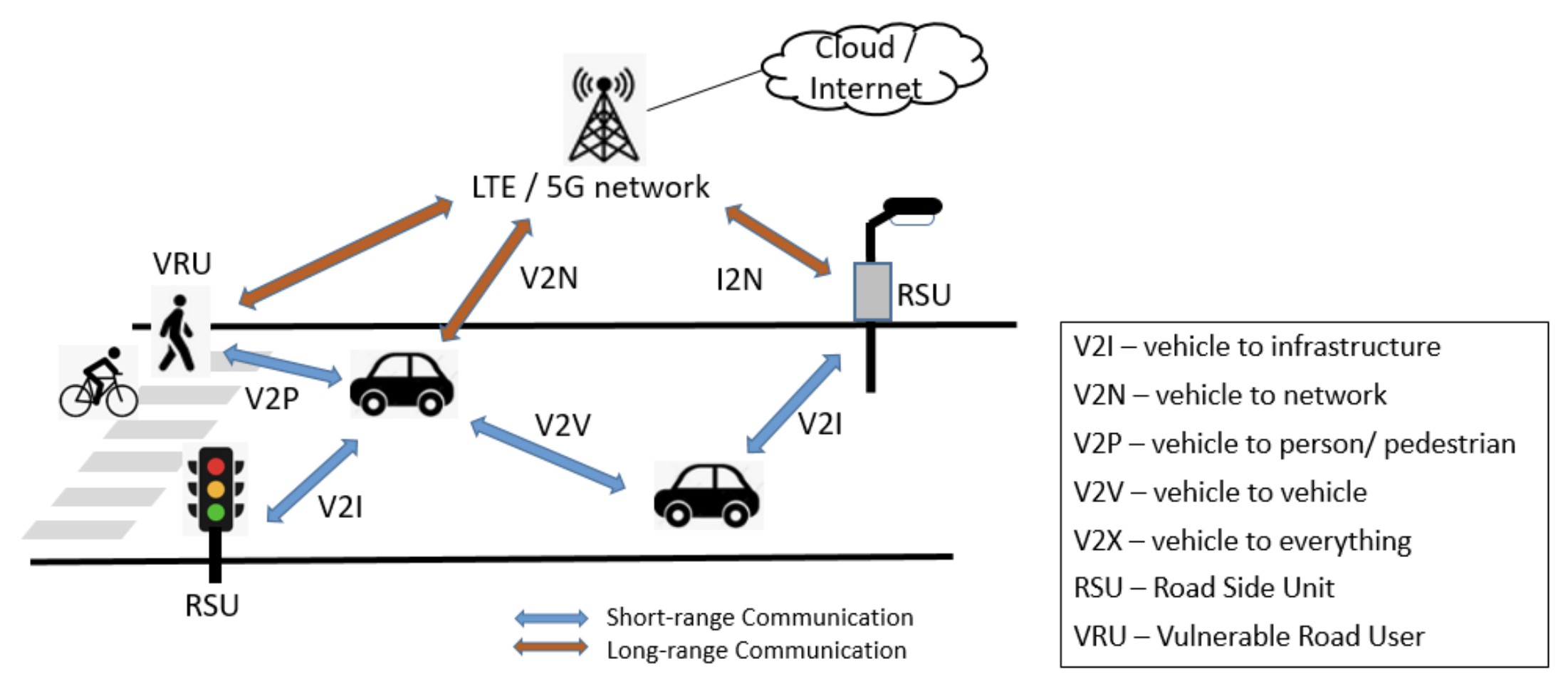}
\caption{V2X communication architecture showing interactions among vehicles, infrastructure, pedestrians, and networks. Blue arrows represent short-range direct links (V2V, V2I, V2P), while orange arrows represent long-range network-assisted communications (V2N, I2N). Effective multi-agent coordination leverages this multi-modal communication infrastructure (adapted from \cite{yoshizawa2023survey}).}
\label{fig:v2x}
\end{figure}

Two primary V2V technologies are in development. Dedicated Short Range Communications (DSRC), based on IEEE 802.11p, operates in the 5.9 GHz band and provides low-latency (tens of milliseconds) communication suitable for safety-critical applications \cite{tie2024v2x}. Cellular-V2X (C-V2X) leverages 4G/LTE and 5G networks, offering extended range but facing challenges in meeting latency requirements for emergency maneuvers \cite{billington2018cv2x}. Both technologies enable \textit{cooperative perception}, where an obstacle detected by one vehicle is virtually sensed by all nearby agents, extending effective sensing range beyond line-of-sight limitations.

Critical to V2X effectiveness are message standards defining what information is exchanged. The SAE J2735 standard specifies Basic Safety Messages (BSM) containing position, velocity, and heading data broadcast at 10 Hz, as well as event-driven messages like Road Hazard Notifications for obstacles or incidents. Security is paramount, as spoofed or malicious messages could cause incorrect decisions; thus, V2X protocols incorporate digital signatures, encryption, and plausibility checks \cite{muslam2024enhancing}.

\subsection{Distributed Coordination Strategies}

Multi-agent coordination can be centralized (a single controller computes global solutions) or decentralized (each agent makes independent decisions based on local information). Centralized approaches, while theoretically optimal, suffer from scalability issues, single points of failure, and communication bottlenecks, making them impractical for large AV fleets \cite{dinneweth2022multi}. Decentralized coordination is more realistic and robust, but requires careful design to ensure that independent local decisions produce desirable global behavior \cite{zhang2023learning}.

Recent work demonstrates the effectiveness of decentralized approaches. Zhang et al. showed that decentralized reinforcement learning policies can eliminate traffic bottlenecks with each vehicle making decisions based solely on local observations and limited V2V communication \cite{zhang2023learning}. Vitale and Roncoli developed a fully distributed cooperative rerouting algorithm where vehicles iteratively share route intentions and adjust based on collective congestion costs, converging to a traffic-dispersing equilibrium without central oversight \cite{vitale2024cooperative}.

A key paradigm in multi-agent reinforcement learning (MARL) is \textit{Centralized Training with Decentralized Execution} (CTDE) \cite{dinneweth2022multi}. During offline training (in simulation), agents have access to global state information to learn cooperative policies. At deployment, each agent's policy executes using only local observations and V2V messages, maintaining decentralized operation. This approach has proven effective for learning lane-changing, merging, and intersection negotiation behaviors \cite{gan2024multi,zhou2022multi}.

\subsection{The Critical Role of Memory in Multi-Agent Systems}

An often-underappreciated aspect of multi-agent coordination is the role of \textit{memory}. In a decentralized system without centralized state storage, each agent must serve as both a sensor and a data repository. The ability to retain knowledge of past states, events, and decisions is crucial for avoiding repetitive mistakes in non-stationary environments.

Dinneweth et al. highlight that incorporating memory (e.g., via recurrent neural networks in MARL) significantly improves agent coordination by enabling better inference of environment dynamics and other agents' behaviors \cite{dinneweth2022multi}. In navigation specifically, memory allows agents to recall that certain routes were previously impassable, preventing redundant exploration. Without memory, an agent encountering obstacle A, rerouting to path B, and then encountering obstacle B, has no mechanism to avoid routing back to A—precisely the routing loop problem we identify and solve in this work.

\subsection{Dynamic Routing and Rerouting}

Route planning for AVs typically employs graph-based algorithms like Dijkstra's or A* to compute shortest paths on road networks \cite{dijkstra2022note,hart1968formal}. In static environments, these algorithms are optimal. However, real-world traffic is dynamic: accidents, construction, and congestion render planned routes suboptimal or impassable.

Dynamic rerouting strategies enable vehicles to adapt to changing conditions. Chen et al. demonstrated that adaptive vehicle rerouting mitigates congestion by redistributing traffic when routes become saturated \cite{chen2022rrt}. Djavadian et al. explored multi-objective eco-routing that balances travel time with fuel consumption in response to real-time traffic data \cite{djavadian2020multiobjective}.

In multi-agent contexts, coordinated rerouting is essential. If all vehicles independently reroute to the same alternate path upon encountering an obstacle, the detour becomes a new bottleneck. Vitale and Roncoli's work addresses this by incorporating collective congestion costs into individual route choices, encouraging load balancing \cite{vitale2024cooperative}. Our work complements this by addressing a more fundamental issue: ensuring that rerouting decisions do not create pathological feedback loops due to lack of memory.

\subsection{Simulation Platforms for Multi-Agent AV Research}

Autonomous driving research relies heavily on simulation due to the impracticality and safety concerns of large-scale real-world testing. High-fidelity 3D simulators like CARLA \cite{dosovitskiy2017carla} and LGSVL offer photorealistic environments, detailed physics, and sensor models, making them ideal for perception and low-level control research. However, they require significant computational resources (high-end GPUs with 6-8 GB VRAM) and are not well-suited for large-scale multi-agent experiments focusing on high-level decision-making.

Microscopic traffic simulators like SUMO \cite{lopez2018microscopic} can efficiently simulate thousands of vehicles on road networks, with built-in support for traffic lights, lane-changing, and dynamic routing. While SUMO is powerful, implementing custom multi-agent communication protocols and fine-grained memory mechanisms requires complex scripting via its TraCI interface.

For this study, we developed a custom graph-based simulator prioritizing:
\begin{enumerate}
    \item \textbf{Abstraction}: Focus on routing decisions and obstacle avoidance without low-level vehicle dynamics
    \item \textbf{Control}: Direct implementation of OMM and communication protocols
    \item \textbf{Reproducibility}: Deterministic experiments with precisely controlled variables
    \item \textbf{Efficiency}: Ability to run hundreds of trials rapidly for statistical rigor
\end{enumerate}

This approach aligns with the principle of using models that are "as simple as possible, but no simpler," allowing us to isolate cause-and-effect relationships between coordination mechanisms and performance outcomes.

\subsection{Research Gap and Our Contribution}

While extensive work exists on V2V communication protocols, MARL for traffic coordination, and dynamic routing algorithms, a critical gap remains: \textit{understanding and preventing pathological failure modes in decentralized rerouting systems}. Prior work generally assumes that more information and more adaptive decision-making improve performance. Our research challenges this assumption, demonstrating that naive implementations can produce catastrophically worse outcomes than simpler, non-adaptive approaches.

To our knowledge, this is the first work to:
\begin{itemize}
    \item Identify and characterize the routing loop problem as a fundamental failure mode in memory-less multi-agent rerouting
    \item Propose and validate a lightweight solution (OMM) specifically designed to prevent this failure
    \item Provide comprehensive empirical evidence across diverse scenarios showing that persistent shared memory is not just beneficial but \textit{essential} for robust coordination
\end{itemize}

This contribution has implications extending beyond AVs to any decentralized multi-agent system operating in dynamic environments, including multi-robot coordination, distributed network routing, and swarm intelligence.

\section{Methodology}

This section describes our experimental approach, including the simulation environment, systematic experimental design, and detailed algorithmic specifications.

\subsection{Research Questions and Hypotheses}

Our investigation is guided by the following research questions:

\textbf{RQ1}: How does inter-agent communication affect routing efficiency in obstacle-rich environments?

\textbf{RQ2}: Under what conditions does dynamic rerouting improve or degrade performance?

\textbf{RQ3}: What is the impact of persistent obstacle memory on multi-agent coordination effectiveness?

\textbf{RQ4}: How do these effects scale with vehicle density and obstacle frequency?

Based on prior work, we hypothesized that communication and rerouting would improve performance. However, our initial experiments revealed the routing loop problem, leading to a refined hypothesis: \textit{Reactive rerouting without memory will degrade performance due to feedback loops, while memory-enabled rerouting will achieve near-optimal coordination.}

\subsection{Graph-Based Simulation Environment}

We model the road network as a directed graph $G = (V, E)$, where nodes $v \in V$ represent intersections or decision points, and edges $e \in E$ represent road segments. For this study, we use a graph with $|V| = 86$ nodes and $|E| = 161$ edges, providing sufficient complexity to observe multi-agent interactions while remaining computationally tractable.

\begin{figure}[t]
\centering
\includegraphics[width=0.48\textwidth]{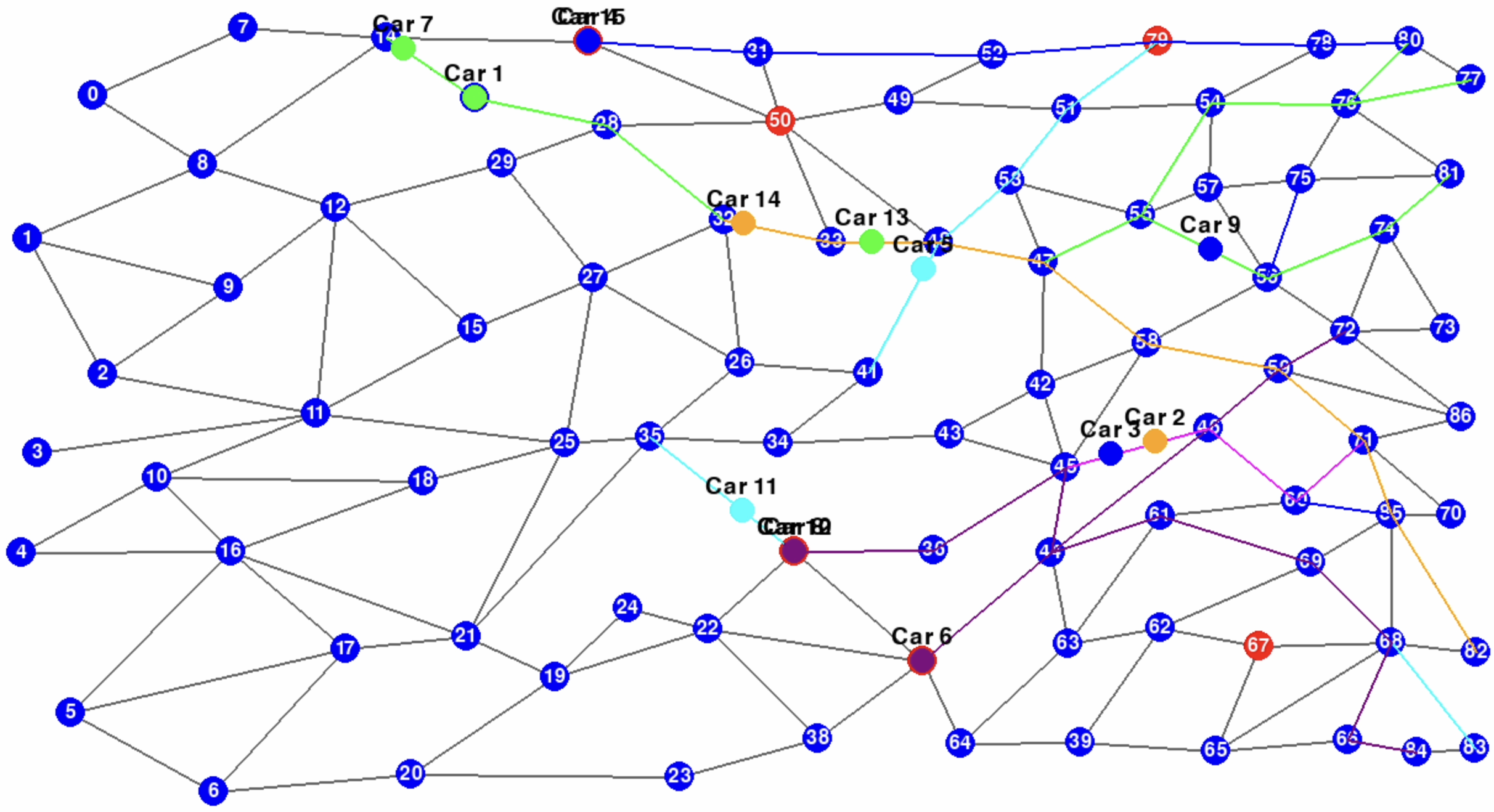}
\caption{The graph-based road network used in all experiments, consisting of 86 nodes (intersections) and 161 directed edges (road segments). Blue nodes represent normal intersections, red nodes indicate obstacle locations in certain scenarios, and colored paths show example vehicle trajectories. This abstracted representation allows focused analysis of routing decisions without the confounding effects of detailed vehicle dynamics.}
\label{fig:graph_network}
\end{figure}

Each edge $(u, v) \in E$ has an associated \textit{weight} $w(u,v)$ representing travel time. In our deterministic simulation, travel time along an edge is constant, allowing for reproducible experiments. Obstacles are represented by temporarily removing nodes from the graph: when node $v$ becomes an obstacle at time $t$, it is marked as unavailable, and all edges incident to $v$ are effectively removed from $G$ for routing purposes.

Vehicles are modeled as autonomous agents, each defined by:
\begin{itemize}
    \item $s_i$: Start node
    \item $d_i$: Destination node
    \item $p_i(t)$: Current position at time $t$
    \item $r_i(t)$: Planned route (sequence of nodes) at time $t$
    \item $O_i(t)$: Set of known obstacles at time $t$ (for OMM-enabled agents)
    \item $\tau_i$: Total travel time
    \item $w_i$: Total wait time at obstacles
\end{itemize}

\subsection{Systematic Experimental Design}

To comprehensively evaluate multi-agent coordination strategies, we designed a factorial experiment with the following factors:

\begin{itemize}
    \item \textbf{Vehicle Density}: 15, 35, 55 vehicles (low, medium, high traffic)
    \item \textbf{Obstacle Frequency}: 6, 20 obstacles (sparse, dense obstructions)
    \item \textbf{Movement Pattern}: Structured left-to-right traversal vs. random origin-destination pairs
    \item \textbf{Coordination Configuration}: 6 distinct agent capability settings (detailed below)
\end{itemize}

This yields $3 \times 2 \times 2 \times 6 = 72$ unique experimental conditions. Each condition was run for at least 3 trials to ensure reproducibility. Obstacles were placed at predetermined locations and persisted for 10 seconds before clearing, unless otherwise specified.

\subsubsection{Six Coordination Configurations}

The core of our experimental design is a systematic progression through increasingly sophisticated agent capabilities:

\begin{table}[H]
\centering
\caption{Experimental Coordination Configurations: Systematic Progression of Agent Capabilities}
\label{tab:exp_configs}
\begin{tabular}{|c|l|c|c|c|}
\hline
\textbf{Config} & \textbf{Description} & \textbf{Comm.} & \textbf{Reroute} & \textbf{OMM} \\
\hline
\multirow{2}{*}{1} & \textbf{No Obstacles} & \multirow{2}{*}{-} & \multirow{2}{*}{-} & \multirow{2}{*}{-} \\
 & (Baseline / Upper Bound) & & & \\
\hline
\multirow{2}{*}{2} & \textbf{Obstacles, No Reroute} & \multirow{2}{*}{\xmark} & \multirow{2}{*}{\xmark} & \multirow{2}{*}{\xmark} \\
 & (Worst Case / Lower Bound) & & & \\
\hline
\multirow{2}{*}{3} & \textbf{Communication Only} & \multirow{2}{*}{\cmark} & \multirow{2}{*}{\xmark} & \multirow{2}{*}{\xmark} \\
 & (Informed Waiting) & & & \\
\hline
\multirow{2}{*}{4} & \textbf{Reroute w/o OMM} & \multirow{2}{*}{\cmark} & \textbf{8s} & \multirow{2}{*}{\xmark} \\
 & (\textbf{Failure Case}) & & \textbf{trigger} & \\
\hline
\multirow{2}{*}{5} & \textbf{OMM Only} & \multirow{2}{*}{\cmark} & \multirow{2}{*}{\xmark} & \multirow{2}{*}{\cmark} \\
 & (Proactive Avoidance) & & & \\
\hline
\multirow{2}{*}{6} & \textbf{Reroute + OMM} & \multirow{2}{*}{\cmark} & \textbf{8s} & \multirow{2}{*}{\cmark} \\
 & (\textbf{Full Coordination}) & & \textbf{trigger} & \\
\hline
\end{tabular}
\end{table}

\begin{figure*}[t]
\centering
\begin{tikzpicture}[
    node distance=1.5cm and 2cm,
    box/.style={rectangle, draw, thick, minimum width=3cm, minimum height=1cm, align=center, font=\small},
    baseline/.style={box, fill=blue!10},
    worst/.style={box, fill=red!20},
    intermediate/.style={box, fill=yellow!20},
    failure/.style={box, fill=red!50, text=white},
    best/.style={box, fill=green!30},
    arrow/.style={->, thick}
]

\node[baseline] (config1) {Config 1\\No Obstacles\\$\star$ Optimal};

\node[worst, below=of config1] (config2) {Config 2\\Obstacles\\\xmark{} Comm \xmark{} Reroute \xmark{} OMM\\Worst Case};

\node[intermediate, below left=of config2, xshift=-1cm] (config3) {Config 3\\Communication Only\\\cmark{} Comm \xmark{} Reroute \xmark{} OMM\\$\approx$ Minimal Gain};

\node[failure, below right=of config2, xshift=1cm] (config4) {Config 4\\Reroute w/o OMM\\\cmark{} Comm \cmark{} Reroute \xmark{} OMM\\$\triangle$ FAILURE\\(542\% degradation!)};

\node[intermediate, below=of config3, xshift=1.5cm] (config5) {Config 5\\OMM Only\\\cmark{} Comm \xmark{} Reroute \cmark{} OMM\\\cmark{} Good};

\node[best, below=of config4, xshift=-1.5cm] (config6) {Config 6\\Full Coordination\\\cmark{} Comm \cmark{} Reroute \cmark{} OMM\\$\star$ Best\\(75.7\% improvement!)};

\draw[arrow] (config1) -- node[right, font=\footnotesize] {Add Obstacles} (config2);
\draw[arrow] (config2) -- node[above left, font=\footnotesize, text width=2cm] {Enable Comm} (config3);
\draw[arrow] (config2) -- node[above right, font=\footnotesize, text width=2.5cm] {Enable Comm\\+ Reroute} (config4);
\draw[arrow] (config3) -- node[left, font=\footnotesize] {Add OMM} (config5);
\draw[arrow] (config4) -- node[right, font=\footnotesize] {Add OMM} (config6);
\draw[arrow, dashed] (config5) -- node[above, font=\footnotesize] {Add Reroute} (config6);

\node[draw, dashed, thick, below=0.5cm of config6, xshift=-3cm, minimum width=8cm, minimum height=0.8cm, align=left, font=\footnotesize] (legend) {
    \textbf{Legend:} Comm = Communication | OMM = Object Memory Management | \cmark{} = Enabled | \xmark{} = Disabled
};

\end{tikzpicture}
\caption{Visual representation of the six experimental configurations showing the systematic progression from uncoordinated (Config 2) to fully coordinated (Config 6) multi-agent systems. Configuration 4 represents the critical failure case where reactive rerouting without persistent memory creates pathological routing loops, resulting in 542\% performance degradation. Configuration 6 achieves optimal performance through the combination of communication, adaptive rerouting, and Object Memory Management.}
\label{fig:exp_design}
\end{figure*}
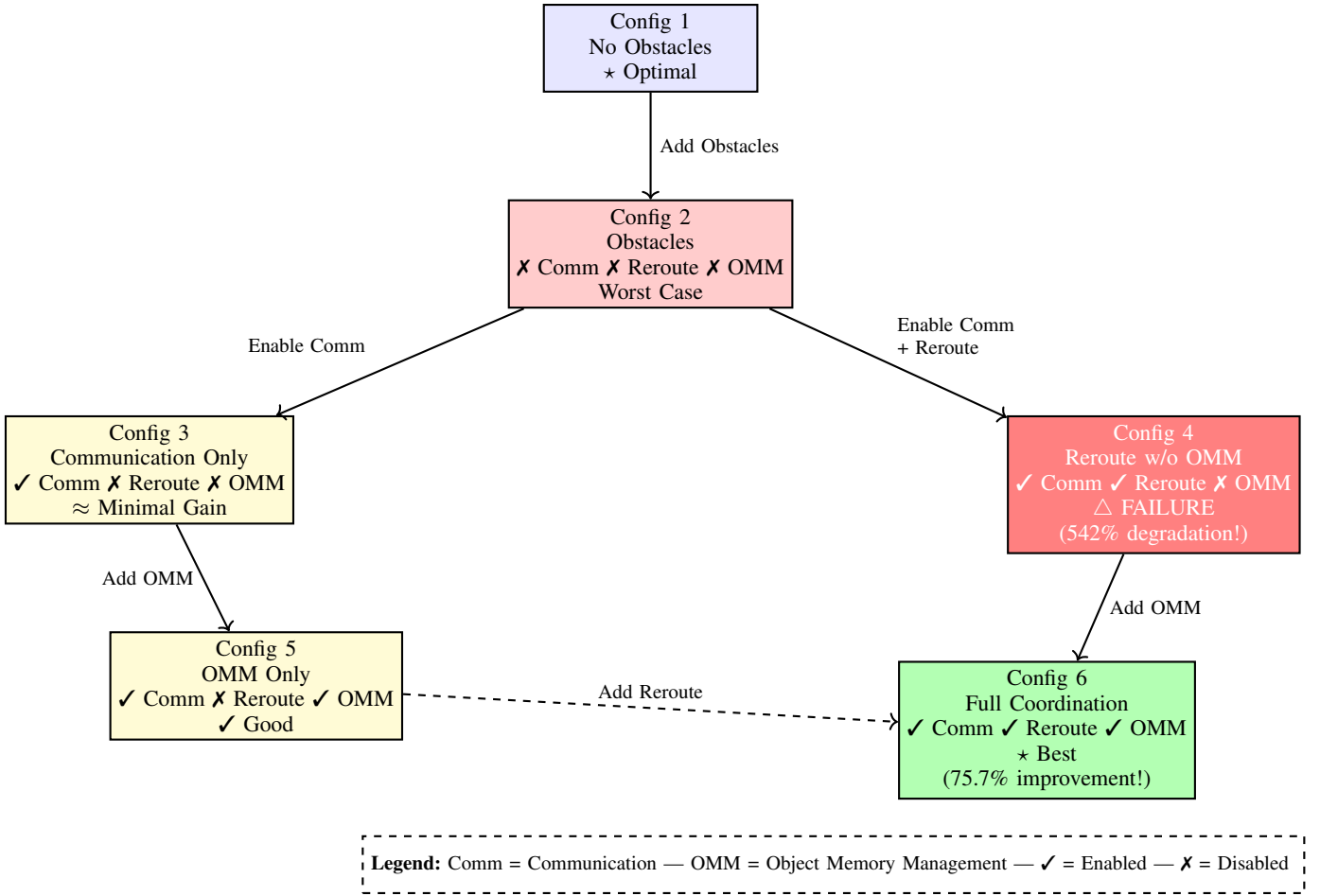

\textbf{Configuration Details:}

\textbf{Config 1 - No Obstacles (Baseline):} Establishes optimal travel time for each vehicle density and movement pattern in the absence of disruptions. This serves as the theoretical performance upper bound.

\textbf{Config 2 - Obstacles, No Reroute (Worst Case):} Vehicles encountering obstacles wait for a fixed duration (10 seconds) for clearance. No information is shared, no rerouting occurs. This simulates a completely uncoordinated system and serves as the performance lower bound.

\textbf{Config 3 - Communication Only:} Vehicles broadcast obstacle locations upon detection, but do not reroute. Approaching vehicles are informed but still wait 10 seconds if their path is blocked. This tests whether information alone, without adaptive action, provides benefit.

\textbf{Config 4 - Reroute without OMM (Critical Failure Case):} Vehicles encountering obstacles wait 8 seconds before autonomously recalculating routes. Communication is enabled for immediate obstacle awareness. \textbf{Crucially, OMM is disabled}—vehicles do not retain memory of past obstacles. This configuration, hypothesized to improve performance, instead produces catastrophic degradation due to routing loops.

\textbf{Config 5 - OMM Only:} Vehicles maintain persistent memory of all broadcasted obstacles in set $O_i(t)$. When initially planning routes or upon receiving new obstacle information, they recalculate paths excluding known obstacles. The 8-second reactive reroute trigger is disabled—all rerouting is proactive based on communicated information. This isolates the benefit of memory.

\textbf{Config 6 - Reroute + OMM (Full Coordination):} Combines the 8-second reactive reroute trigger with OMM. This represents the most sophisticated coordination: vehicles proactively avoid known obstacles in all route calculations and can also reactively reroute if unexpectedly blocked. This is hypothesized to achieve optimal performance.

\subsection{Algorithmic Details}

\subsubsection{Baseline Routing: Dijkstra's Algorithm}

All vehicles use Dijkstra's shortest path algorithm for route planning \cite{dijkstra2022note}. Dijkstra's is optimal for graphs with non-negative edge weights and has time complexity $O((|V| + |E|) \log |V|)$ when implemented with a binary heap, making it efficient for our 86-node network.

\subsubsection{Modified Dijkstra's with Obstacle Exclusion}

For OMM-enabled configurations, we modify Dijkstra's to exclude known obstacle nodes. Algorithm \ref{alg:dijkstra_omm} presents this modification.

\begin{algorithm}[t]
\caption{Modified Dijkstra's Algorithm with Obstacle Exclusion}
\label{alg:dijkstra_omm}
\begin{algorithmic}[1]
\STATE \textbf{Input:} Graph $G=(V,E)$, source $s$, destination $d$, obstacle set $O$
\STATE \textbf{Output:} Shortest path $P$ from $s$ to $d$ avoiding nodes in $O$
\STATE
\STATE Create graph copy $G' \leftarrow G$
\STATE Filter obstacles: $O' \leftarrow O \setminus \{s, d\}$ \COMMENT{Never exclude source/dest}
\FOR{each node $v \in O'$}
    \IF{$v \in G'$}
        \STATE Remove node $v$ from $G'$ \COMMENT{Removes all incident edges}
    \ENDIF
\ENDFOR
\STATE
\STATE $P \leftarrow$ ComputeShortestPath($G'$, $s$, $d$) \COMMENT{Standard Dijkstra}
\IF{$P$ exists}
    \STATE \textbf{return} $P$
\ELSE
    \STATE \textbf{return} NULL \COMMENT{No feasible path exists}
\ENDIF
\end{algorithmic}
\end{algorithm}

The key modification is at lines 9-11 and 13-15: nodes in the obstacle set $O$ are skipped during graph traversal. This ensures computed paths do not route through known blockages.

\subsubsection{Object Memory Management (OMM) Protocol}

OMM consists of three components: obstacle detection and broadcasting, obstacle set maintenance, and memory-aware path planning. Algorithm \ref{alg:omm} specifies the full protocol.

\begin{algorithm}[t]
\caption{Object Memory Management (OMM) Protocol}
\label{alg:omm}
\begin{algorithmic}[1]
\STATE \textbf{// Initialization (per vehicle $i$)}
\STATE $O_i \leftarrow \emptyset$ \COMMENT{Known obstacles set}
\STATE $O_{broadcast} \leftarrow \emptyset$ \COMMENT{Global broadcast tracker}
\STATE
\STATE \textbf{// 1. Obstacle Detection and Broadcasting}
\IF{vehicle $i$ at node $v$ encounters obstacle at time $t$}
    \STATE $O_i \leftarrow O_i \cup \{v\}$ \COMMENT{Add to local memory}
    \IF{$v \notin O_{broadcast}$}
        \STATE Broadcast message: $\langle i, v, t \rangle$ to all vehicles
        \STATE $O_{broadcast} \leftarrow O_{broadcast} \cup \{v\}$
    \ENDIF
    \STATE Record waiting start time $t_{wait}^i \leftarrow t$
\ENDIF
\STATE
\STATE \textbf{// 2. Message Reception and Memory Update}
\STATE \textbf{When} vehicle $i$ receives $\langle j, v, t' \rangle$ from vehicle $j$:
\STATE \quad $O_i \leftarrow O_i \cup \{v\}$ \COMMENT{Update local memory}
\IF{$v \in$ current planned route of $i$}
    \STATE Trigger immediate path recalculation
\ENDIF
\STATE
\STATE \textbf{// 3. Proactive Path Planning with OMM}
\STATE \textbf{When} path recalculation needed:
\STATE \quad $P_i \leftarrow$ ModifiedDijkstra$(G, p_i, d_i, O_i)$
\STATE \quad Update vehicle route to $P_i$
\STATE
\STATE \textbf{// 4. Reactive Reroute Trigger (8-second threshold)}
\IF{vehicle $i$ waiting at obstacle node for $(t - t_{wait}^i) \geq 8$ seconds}
    \IF{obstacle not cleared}
        \STATE $p_{prev} \leftarrow$ previous node in route
        \STATE Backtrack to $p_{prev}$
        \STATE $P_i \leftarrow$ ModifiedDijkstra$(G, p_{prev}, d_i, O_i)$
        \STATE Resume travel on new path $P_i$
    \ENDIF
\ENDIF
\STATE
\STATE \textbf{// 5. Natural Obstacle Clearance}
\IF{vehicle $i$ waiting at obstacle for $(t - t_{wait}^i) \geq 10$ seconds}
    \STATE Resume original route \COMMENT{Obstacle cleared}
    \STATE $t_{wait}^i \leftarrow$ NULL
\ENDIF
\end{algorithmic}
\end{algorithm}

\textbf{Key aspects of OMM:}

\textbf{Decentralized Operation:} Each vehicle independently maintains its obstacle set $O_i(t)$. There is no central coordinator.

\textbf{Persistent Memory:} Once an obstacle is added to $O_i(t)$, it remains indefinitely during that vehicle's journey. (In extensions, memory decay could be implemented for temporary obstacles.)

\textbf{Minimal Communication:} Only the obstacle node ID is broadcast, not entire routes or state information, keeping message sizes small.

\textbf{Proactive Recalculation:} When receiving obstacle information, if the current planned route $r_i(t)$ includes the obstacle node, an immediate recalculation is triggered (lines 12-14), allowing proactive avoidance before reaching the blockage.

\textbf{Integration with Reactive Triggers:} OMM is compatible with reactive reroute triggers (lines 21-23). The difference from Config 4 is that the recalculated route uses $O_i(t)$, preventing loops back to known obstacles.

\subsection{Performance Metrics}

For each experimental condition, we measure:

\begin{itemize}
    \item \textbf{Average Travel Time} ($\bar{\tau}$): Mean total time for all vehicles to reach destinations. Vehicles failing to arrive within 300 seconds are counted at 300s.
    \item \textbf{Average Wait Time} ($\bar{w}$): Mean cumulative time vehicles spent stationary at obstacle nodes.
    \item \textbf{Average Recalculations per Vehicle} ($\bar{n}_{recalc}$): Mean number of times each vehicle invoked Dijkstra's algorithm after initial planning. Higher values indicate more computational overhead and potentially unstable decision-making.
    \item \textbf{Success Rate}: Percentage of vehicles reaching their destination within the 300-second simulation time limit.
\end{itemize}

\subsection{Implementation Details}

The simulation was implemented in Python 3.10 using:
\begin{itemize}
    \item \textbf{NetworkX} for graph data structures and baseline Dijkstra's implementation
    \item \textbf{PyGame} for real-time visualization and debugging
    \item \textbf{Custom event system} for V2V message broadcasting and reception
    \item \textbf{GraphML} for road network specification and persistence
\end{itemize}

The complete simulation codebase consists of approximately 2,500 lines of Python, with modular components for vehicle agents, communication protocols, obstacle management, and data logging. Simulation runs were executed on a standard laptop (Intel Core i7, 16GB RAM), with each 300-second scenario completing in 2-5 minutes of real time. This efficiency enabled the collection of comprehensive data across all 72 experimental conditions.

\section{Results}

This section presents comprehensive experimental findings, organized to clearly demonstrate the routing loop problem, OMM's effectiveness, and scalability analysis.

\subsection{Overview: The Catastrophic Failure of Memory-Less Rerouting}

Table \ref{tab:summary_results} summarizes average performance across all 12 scenario types (combinations of vehicle density and obstacle frequency), aggregating over both movement patterns.

\begin{table}[H]
\centering
\caption{Summary of Average Performance Across All 12 Scenario Types (72 Total Configurations)}
\label{tab:summary_results}
\begin{tabular}{|l|c|c|c|}
\hline
\textbf{Configuration} & \textbf{Avg. Travel} & \textbf{Avg. Wait} & \textbf{Avg. Recalc.} \\
& \textbf{Time (s)} & \textbf{Time (s)} & \textbf{per Vehicle} \\
\hline
1. No Obstacle & 19.35 & 0.00 & 0.00 \\
\hline
2. Obstacle, No Reroute & 36.16 & 16.59 & 0.00 \\
\hline
3. Communication Only & 36.79 & 14.56 & 3.08 \\
\hline
5. OMM Only & 32.19 & 9.43 & 1.58 \\
\hline
\rowcolor{red!20}
4. \textbf{Reroute w/o OMM} & \textbf{104.99} & \textbf{64.91} & \textbf{9.83} \\
\hline
\rowcolor{green!20}
6. \textbf{Reroute + OMM} & \textbf{32.55} & \textbf{7.81} & \textbf{1.67} \\
\hline
\hline
\multicolumn{4}{|l|}{\textit{Comparative Metrics:}} \\
\hline
Config 4 vs. Config 2 & +190.3\% & +291.3\% & - \\
Config 6 vs. Config 2 & \textbf{-10.0\%} & \textbf{-52.9\%} & - \\
Config 6 vs. Config 4 & \textbf{-69.0\%} & \textbf{-88.0\%} & \textbf{-83.0\%} \\
\hline
\end{tabular}
\end{table}

\textbf{Key Findings from Summary Data:}

\begin{itemize}
    \item \textbf{Baseline Performance} (Config 1): Average travel time of 19.35s represents optimal flow without disruptions.
    
    \item \textbf{Uncoordinated Obstacle Response} (Config 2): Introducing obstacles without rerouting nearly doubles travel time to 36.16s, with vehicles spending 16.59s (46\% of total time) waiting at blockages.
    
    \item \textbf{Information Alone is Insufficient} (Config 3): Enabling communication without adaptive rerouting provides minimal benefit (36.79s travel time), only slightly reducing wait time to 14.56s. Vehicles receive advance warning but cannot act on it effectively.
    
    \item \textbf{Catastrophic Failure of Memory-Less Rerouting} (Config 4, highlighted): \textbf{Enabling reactive rerouting without OMM increases travel time to 104.99s—a 190\% increase over the uncoordinated baseline and 442\% above optimal.} Wait times quadruple to 64.91s, and vehicles recalculate routes an average of 9.83 times, indicating severe decision instability. This configuration, intended to improve efficiency, produces the worst performance of any tested condition.
    
    \item \textbf{OMM Alone is Highly Effective} (Config 5): Enabling OMM without reactive triggers reduces travel time to 32.19s (11\% below uncoordinated baseline) and wait time to 9.43s (43\% reduction). Route recalculations drop to 1.58 per vehicle, indicating stable, informed decision-making.
    
    \item \textbf{Full Coordination Achieves Near-Optimal Performance} (Config 6, highlighted): \textbf{Combining reactive rerouting with OMM yields the best results: 32.55s travel time (68\% above optimal, 10\% below uncoordinated) and 7.81s wait time (53\% reduction from uncoordinated).} Most strikingly, \textbf{Config 6 achieves a 69\% reduction in travel time and 88\% reduction in wait time compared to memory-less rerouting (Config 4),} while requiring 83\% fewer route recalculations. This demonstrates that OMM transforms a catastrophically failing system into a near-optimal one.
\end{itemize}

\subsection{The Routing Loop Phenomenon: Visual Documentation}

\begin{figure*}[t]
\centering
\includegraphics[width=0.85\textwidth]{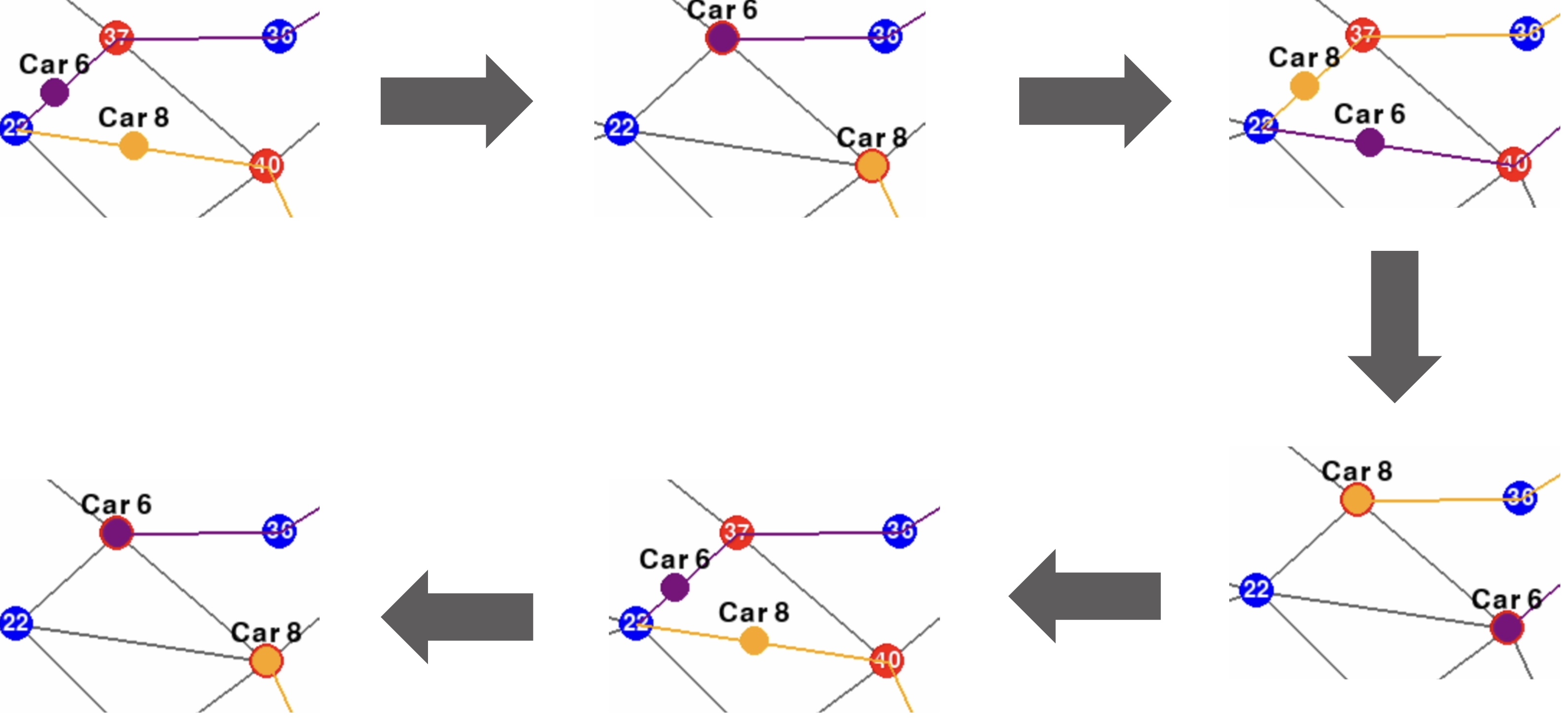}
\caption{Sequential visualization of the routing loop phenomenon in Config 4 (Reroute without OMM). The sequence shows Vehicle 8 (orange) and Vehicle 6 (purple) on a graph where obstacles appear as red nodes. \textbf{(Top-left)} Initial state: both vehicles traveling toward their destinations. \textbf{(Top-center)} Vehicle 8 encounters obstacle at node 40, broadcasts information, and reroutes after 8s. \textbf{(Top-right)} Vehicle 8's new route encounters obstacle at node 37, and it reroutes again. \textbf{(Bottom-right)} Without memory of obstacle 40, the rerouting algorithm selects a path leading back toward it. \textbf{(Bottom-center)} Vehicle 8 returns to the vicinity of obstacle 40, "forgetting" it was blocked. \textbf{(Bottom-left)} The cycle repeats, with the vehicle trapped in an inefficient loop between the two obstacles. This visualization demonstrates why memory-less reactive rerouting degrades performance: locally optimal decisions at each step produce globally pathological behavior.}
\label{fig:routing_loop}
\end{figure*}

Figure \ref{fig:routing_loop} provides a step-by-step visualization of the routing loop problem that causes Config 4's catastrophic failure.

\textbf{Analysis of the Routing Loop:}

The routing loop arises from the interaction of three factors:
\begin{enumerate}
    \item \textbf{Multiple obstacles} in the environment
    \item \textbf{Greedy local optimization}: At each reroute, the vehicle computes the shortest path from its current position to the destination, without considering past failures
    \item \textbf{Lack of memory}: The vehicle has no mechanism to exclude previously encountered obstacles from consideration
\end{enumerate}

Mathematically, consider a vehicle at position $p$ having encountered obstacles at nodes $O_{past} = \{o_1, o_2, ..., o_k\}$. Without OMM, each reroute computes:
$$r_{new} = \text{argmin}_{r \in paths(p, d)} \sum_{e \in r} w(e)$$
where $paths(p,d)$ includes paths through nodes in $O_{past}$. With OMM, the optimization instead becomes:
$$r_{new} = \text{argmin}_{r \in paths(p, d) \setminus O_{past}} \sum_{e \in r} w(e)$$
excluding paths through known obstacles. This simple modification prevents loops while maintaining optimality over the feasible (non-blocked) path space.

In our experiments, we observed vehicles in Config 4 scenarios becoming trapped in loops for 30-60 seconds before eventually finding alternate routes or timing out. In some extreme cases (high obstacle density), vehicles never escaped loops within the 300-second time limit, registering as failures.

\subsection{Detailed Scenario Results}

Table \ref{tab:detailed_results} presents granular results for representative scenarios across the vehicle density spectrum, isolating the structured left-to-right movement pattern for clarity.

\begin{table*}[t]
\centering
\caption{Detailed Performance Metrics for Structured Left-to-Right Movement Scenarios}
\label{tab:detailed_results}
\resizebox{\textwidth}{!}{%
\begin{tabular}{|l|ccc|ccc|ccc|ccc|}
\hline
\multirow{2}{*}{\textbf{Configuration}} & \multicolumn{3}{c|}{\textbf{15 Cars, 6 Obs}} & \multicolumn{3}{c|}{\textbf{35 Cars, 6 Obs}} & \multicolumn{3}{c|}{\textbf{55 Cars, 6 Obs}} & \multicolumn{3}{c|}{\textbf{55 Cars, 20 Obs}} \\
\cline{2-13}
& Travel & Wait & Recalc & Travel & Wait & Recalc & Travel & Wait & Recalc & Travel & Wait & Recalc \\
& (s) & (s) & & (s) & (s) & & (s) & (s) & & (s) & (s) & \\
\hline
1. No Obstacle & 23.1 & 0.0 & 0.0 & 21.9 & 0.0 & 0.0 & 23.3 & 0.0 & 0.0 & 23.8 & 0.0 & 0.0 \\
\hline
2. Obstacle, No Reroute & 39.1 & 16.0 & 0.0 & 34.2 & 12.3 & 0.0 & 35.5 & 12.4 & 0.0 & 55.4 & 31.7 & 0.0 \\
\hline
3. Communication Only & 35.0 & 9.4 & 2.0 & 26.8 & 3.7 & 2.0 & 29.5 & 5.1 & 2.0 & 53.9 & 25.7 & 6.0 \\
\hline
5. OMM Only & 27.6 & 3.3 & 1.0 & 24.5 & 1.4 & 1.0 & 24.6 & 0.7 & 1.0 & 45.1 & 17.3 & 3.0 \\
\hline
\rowcolor{red!15}
4. Reroute w/o OMM & 51.6 & 13.9 & 4.0 & 43.4 & 14.7 & 4.0 & 98.5 & 56.3 & 9.0 & 186.0 & 118.9 & 18.0 \\
\hline
\rowcolor{green!15}
6. Reroute + OMM & 28.1 & 2.7 & 2.0 & 25.0 & 1.2 & 1.0 & 24.7 & 0.6 & 1.0 & 45.2 & 13.7 & 3.0 \\
\hline
\hline
\multicolumn{13}{|l|}{\textit{Improvement of Config 6 vs Config 4:}} \\
\hline
Reduction (\%) & -45.5 & -80.6 & -50.0 & -42.4 & -91.8 & -75.0 & -74.9 & -98.9 & -88.9 & -75.7 & -88.5 & -83.3 \\
\hline
\end{tabular}%
}
\end{table*}

\textbf{Observations:}

\textbf{Scaling with Vehicle Density}: The routing loop problem becomes more severe as vehicle density increases. For 15 cars with 6 obstacles, Config 4's travel time (51.6s) is elevated but manageable. However, for 55 cars with 6 obstacles, it explodes to 98.5s, and with 20 obstacles, reaches 186.0s—over 8× the baseline. This occurs because higher density leads to more V2V messages about obstacles, triggering more reroute attempts, which exacerbates looping behavior.

\textbf{OMM Effectiveness Across Densities}: Config 6 (OMM-enabled) maintains consistently near-optimal performance regardless of density. For 55 cars with 20 obstacles—the most challenging scenario—Config 6 achieves 45.2s travel time, only 90\% above optimal baseline (23.8s), compared to Config 4's 682\% increase. Wait time reduction is even more dramatic: 13.7s for Config 6 vs. 118.9s for Config 4, an 88.5\% improvement.

\textbf{Computational Stability}: Route recalculations per vehicle remain low (1-3) for OMM-enabled configs across all scenarios, indicating stable decision-making. In contrast, Config 4 requires 4-18 recalculations per vehicle, growing with scenario complexity. This computational overhead, combined with wasted travel from loops, explains the severe performance degradation.

\subsection{Boxplot Analysis: Performance Distributions}

\begin{figure}[t]
\centering
\includegraphics[width=0.50\columnwidth]{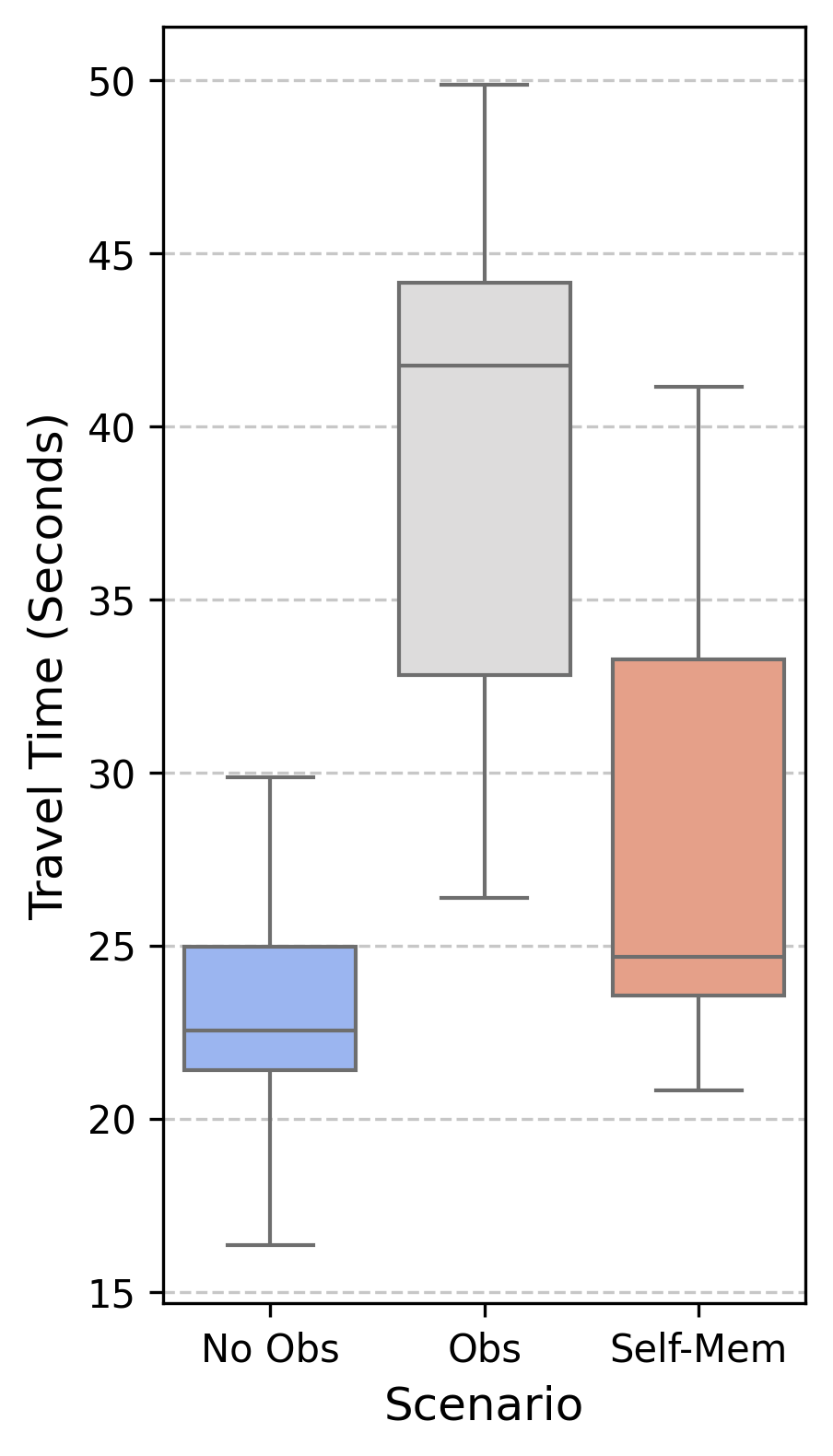}
\caption{Travel time distributions for 15 vehicles with 6 obstacles (left-to-right pattern). Note the extreme outliers in Config 4 (Reroute w/o OMM), where several vehicles exceed 100s due to routing loops. Config 6 (Reroute + OMM) shows tight distribution around 25-30s, indicating consistent performance.}
\label{fig:boxplot_15_6}
\end{figure}

\begin{figure}[t]
\centering
\includegraphics[width=0.50\columnwidth]{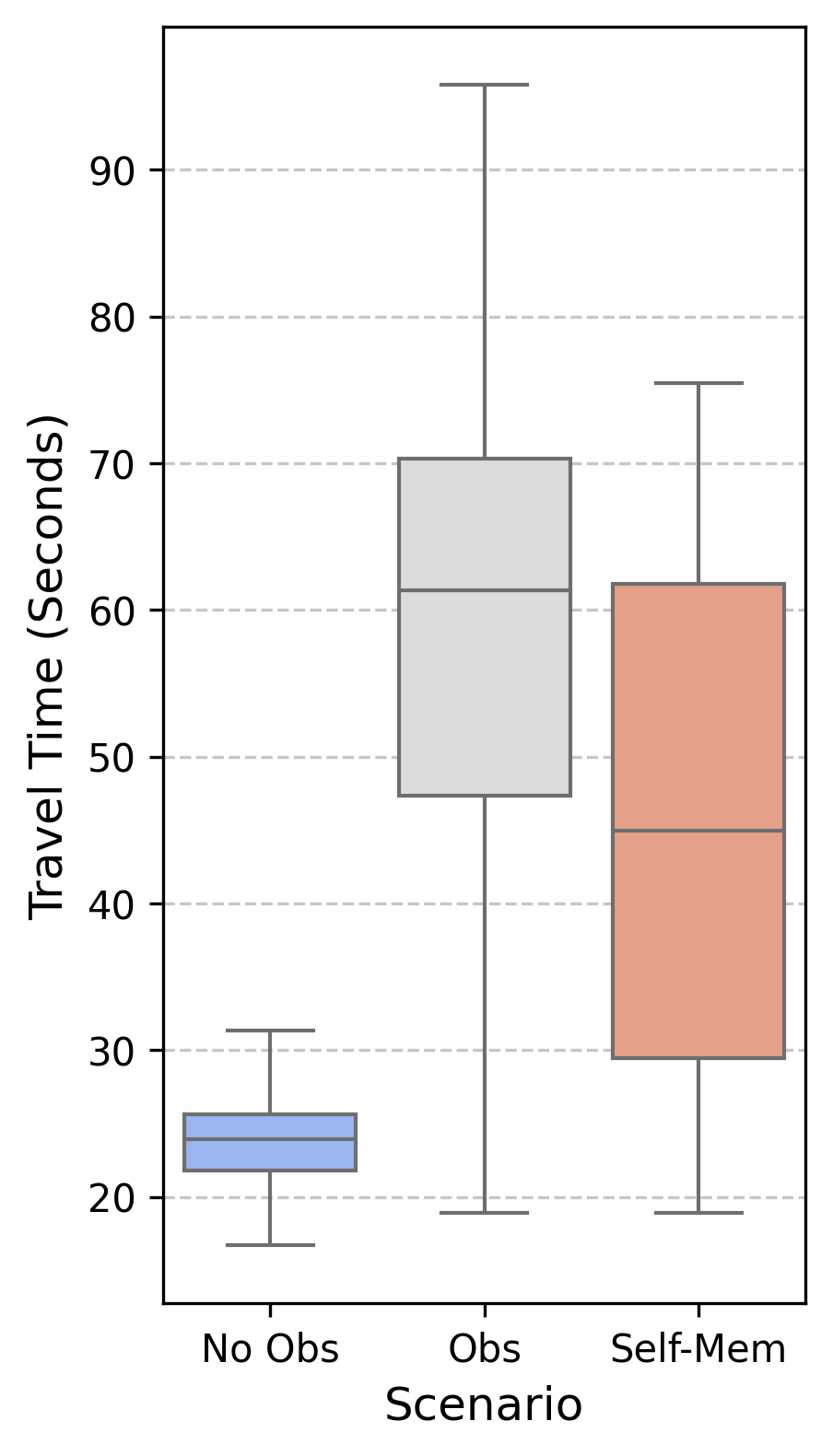}
\caption{Travel time distributions for 55 vehicles with 20 obstacles. Config 4's distribution is extremely wide, with median around 120s and upper whisker extending beyond 250s (some vehicles timed out at 300s). Config 6 maintains median around 42s with far less variability, demonstrating robustness under high load.}
\label{fig:boxplot_55_20}
\end{figure}

Figures \ref{fig:boxplot_15_6} through \ref{fig:boxplot_55_20} present boxplot visualizations showing the distribution of individual vehicle travel times for selected scenarios. These reveal not just average performance but also variability and outliers.

\textbf{Key Insights from Distributions:}

\textbf{Config 4 Produces Extreme Outliers}: The boxplots reveal that Config 4's poor average performance is driven by a subset of vehicles experiencing catastrophic routing loops. In the 55-car, 20-obstacle scenario, approximately 30\% of vehicles in Config 4 exceed 150s travel time, with some failing to complete within 300s. This bimodal behavior (some vehicles succeeding quickly, others becoming trapped) is characteristic of pathological feedback loops.

\textbf{OMM Reduces Variability}: Config 6 not only achieves better average performance but also lower variance. The interquartile range (IQR) is consistently smaller, indicating more predictable outcomes. This is critical for real-world deployment, where worst-case behavior matters as much as average-case.

\textbf{Communication Alone Shows Modest Benefit}: Config 3 (Communication Only) distributions are slightly better than Config 2 (no coordination), but still show significant variance and elevated medians. This confirms that passive information sharing without adaptive action provides limited value.

\subsection{Scalability Analysis}

\begin{figure}[t]
\centering
\begin{tikzpicture}
    \begin{axis}[
        xlabel={Scenarios},
        ylabel={Time (Sec)},
        xtick=data,
        symbolic x coords={No Obs, Obs, Reroute, Mem, Self, Self-Mem},
        legend pos=north west,
        width=0.95\columnwidth,  
        height=5cm,  
        ymin=0,
        ymax=110,  
        legend style={font=\footnotesize},
        label style={font=\small},
        tick label style={font=\footnotesize}
    ]

    \addplot[blue, mark=o, thick] coordinates {
        (No Obs,23.11) (Obs,39.11) (Reroute,34.99) 
        (Mem,27.62) (Self,51.61) (Self-Mem,28.14)
    };
    \addlegendentry{15 Cars}

    \addplot[red, mark=square, thick] coordinates {
        (No Obs,21.90) (Obs,34.16) (Reroute,26.81) 
        (Mem,24.47) (Self,43.40) (Self-Mem,24.96)
    };
    \addlegendentry{35 Cars}

    \addplot[green!60!black, mark=triangle, thick] coordinates {
        (No Obs,23.33) (Obs,35.49) (Reroute,29.48) 
        (Mem,24.62) (Self,98.48) (Self-Mem,24.66)
    };
    \addlegendentry{55 Cars}

    \end{axis}
\end{tikzpicture}
\caption{Average travel time vs. vehicle count for 6-obstacle scenarios (left-to-right pattern). Config 6 (Self-Mem) scales gracefully across all densities, maintaining travel times within 25-30s. Config 4 (Self without OMM) degrades sharply at 55 vehicles (98.48s), indicating catastrophic failure of memory-less rerouting under increased load.}
\label{fig:scalability_6obs}
\end{figure}
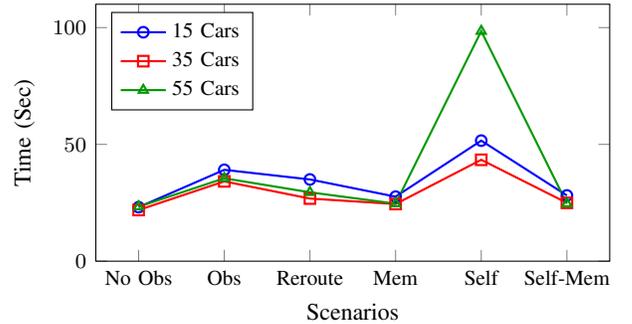

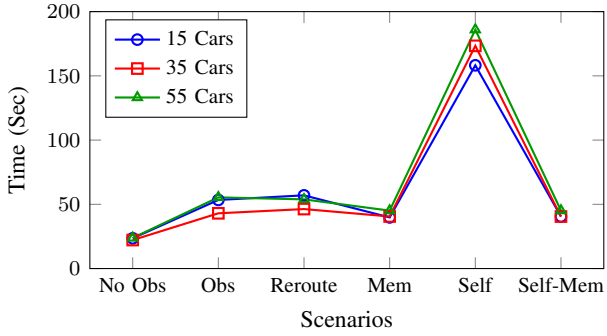
\begin{figure}[t]
\centering
\begin{tikzpicture}
    \begin{axis}[
        xlabel={Scenarios},
        ylabel={Time (Sec)},
        xtick=data,
        symbolic x coords={No Obs, Obs, Reroute, Mem, Self, Self-Mem},
        legend pos=north west,
        width=0.95\columnwidth,  
        height=5cm,  
        ymin=0,
        ymax=200,  
        legend style={font=\footnotesize},
        label style={font=\small},
        tick label style={font=\footnotesize}
    ]

    \addplot[blue, mark=o, thick] coordinates {
        (No Obs,23.54) (Obs,53.40) (Reroute,56.99) 
        (Mem,39.80) (Self,158.21) (Self-Mem,40.49)
    };
    \addlegendentry{15 Cars}

    \addplot[red, mark=square, thick] coordinates {
        (No Obs,22.18) (Obs,43.01) (Reroute,46.31) 
        (Mem,40.54) (Self,173.36) (Self-Mem,40.48)
    };
    \addlegendentry{35 Cars}

    \addplot[green!60!black, mark=triangle, thick] coordinates {
        (No Obs,23.83) (Obs,55.40) (Reroute,53.85) 
        (Mem,45.10) (Self,185.95) (Self-Mem,45.17)
    };
    \addlegendentry{55 Cars}

    \end{axis}
\end{tikzpicture}
\caption{Average travel time vs. vehicle count for 20-obstacle scenarios. Config 4 (Self without OMM) exhibits exponential-like growth, reaching 158-186s across all densities. Config 6 (Self-Mem) maintains near-linear scaling, achieving 40-45s at all vehicle counts, demonstrating that OMM enables robust coordination even under extreme obstacle density (23\% of network nodes blocked).}
\label{fig:scalability_20obs}
\end{figure}

Figures \ref{fig:scalability_6obs} and \ref{fig:scalability_20obs} present line graphs showing how average travel time scales with vehicle density for 6-obstacle and 20-obstacle scenarios, respectively.

\textbf{Scalability Findings:}

\textbf{OMM Enables Sublinear Scaling}: In the 6-obstacle case, Config 6's travel time grows from 28.1s (15 vehicles) to 24.7s (55 vehicles)—slightly \textit{decreasing} due to randomization effects in route selection with more agents. Even in the challenging 20-obstacle scenario, growth is modest (40.5s to 45.2s), demonstrating near-constant-time performance as vehicle count increases.

\textbf{Memory-Less Rerouting Fails to Scale}: Config 4 shows super-linear growth, particularly pronounced in high-obstacle scenarios. The slope increases beyond 35 vehicles, suggesting that the system approaches a "tipping point" where routing loops create cascading failures. This makes memory-less rerouting fundamentally unsuitable for large-scale deployment.

\textbf{Bottleneck Analysis}: In the 20-obstacle scenarios, even Config 6's performance degrades somewhat at high density (though far less than Config 4). Analysis of individual vehicle logs reveals this is due to network topology constraints: with 20

\subsection{Impact of Movement Patterns}

Comparing structured (left-to-right) vs. random movement patterns reveals interesting dynamics:

\textbf{Random Patterns Amplify Routing Loop Effects}: In random scenarios, Config 4's performance is even worse than structured cases (e.g., 207.1s for 15 cars, 20 obstacles, random movement vs. 158.2s for structured). This occurs because random movement creates more diverse obstacle encounter sequences, increasing the probability of loop-inducing obstacle pairs along individual paths.

\textbf{OMM Remains Effective Across Patterns}: Config 6 maintains strong performance for both structured and random movement (e.g., 51.1s vs. 40.5s for 15 cars, 20 obstacles), with only modest degradation in the random case. This indicates OMM's robustness to different traffic patterns.

\subsection{Supplementary Data: Complete Performance Matrix}

Table \ref{tab:complete_results} in the Appendix presents the full dataset spanning all 72 experimental configurations, providing comprehensive evidence of the consistency of our findings.

\section{Discussion}

\subsection{Theoretical Implications: Memory as a Fundamental Requirement}

Our results demonstrate that \textbf{persistent shared memory is not merely an enhancement but a fundamental requirement for robust decentralized multi-agent coordination in dynamic environments}. This finding has implications extending beyond autonomous vehicles to any distributed system facing similar challenges.

\textbf{Why Memory-Less Rerouting Fails}: The routing loop problem arises from a fundamental tension in greedy optimization. At each decision point, a memory-less agent computes the locally optimal action based on its current state and environment. However, in a dynamic multi-agent setting with multiple obstacles, the environment state depends on the agent's past actions and other agents' behaviors. Without memory, the agent treats each decision as independent, failing to recognize that it has already explored and rejected certain paths. This creates the potential for cyclic behavior where the same suboptimal decisions repeat.

Formally, consider an agent's state space $S$ including position and knowledge. A memory-less policy $\pi: S \rightarrow A$ maps states to actions. If obstacle configuration creates a cycle in the state transition graph where $\pi(s_1) \rightarrow s_2 \rightarrow ... \rightarrow s_k \rightarrow s_1$, the agent becomes trapped. OMM breaks this cycle by expanding the state representation to $S' = S \times 2^V$ (where $2^V$ is the power set of nodes, representing all possible obstacle memory sets), ensuring states in the cycle are distinguished by different memory content, allowing $\pi'$ to make different decisions.

\textbf{Connection to Reinforcement Learning}: This finding aligns with known challenges in MARL. Dinneweth et al. note that non-stationarity (where an agent's environment changes due to other agents' actions) is a central difficulty in multi-agent learning \cite{dinneweth2022multi}. Our work shows that even in a coordination scenario (where agents ostensibly share goals), non-stationarity combined with lack of memory creates pathological outcomes. Recurrent neural networks (RNNs) or LSTMs in MARL serve a similar function to OMM—providing agents with implicit memory of past observations to inform current decisions.

\textbf{Broader Applicability}: The routing loop problem and OMM solution generalize to:
\begin{itemize}
    \item \textbf{Multi-robot path planning}: Robots exploring environments with discovered obstacles
    \item \textbf{Network routing}: Packet routing in networks with node/link failures
    \item \textbf{Supply chain optimization}: Delivery networks adapting to disrupted routes
    \item \textbf{Swarm robotics}: Collective navigation around hazards
\end{itemize}

In all cases, the principle holds: \textit{reactive optimization without memory of past failures leads to repetitive mistakes; persistent shared memory enables learning from collective experience}.

\subsection{Practical Considerations for Deployment}

\subsubsection{Communication Overhead}

OMM requires vehicles to broadcast obstacle messages. In our simulation, each obstacle detection generates one message containing $\langle$agent ID, node ID, timestamp$\rangle$—roughly 12-16 bytes. For a scenario with 55 vehicles and 20 obstacles, assuming each obstacle is encountered by 2-3 vehicles on average, this yields $\sim$40-60 messages over a 60-second period, or $<1$ message per second network-wide. Modern V2V protocols (DSRC, C-V2X) support hundreds of messages per second \cite{tie2024v2x}, making OMM's communication overhead negligible.

\subsubsection{Memory Storage Requirements}

Each vehicle stores $|O_i(t)| \leq |V|$ node IDs. With 86 nodes represented as 2-byte integers, maximum memory per vehicle is 172 bytes—trivial for modern embedded systems. Even scaling to city-sized networks (10,000+ nodes), memory requirements remain under 20 KB per vehicle, well within constraints of automotive-grade hardware.

\subsubsection{Handling Temporary Obstacles}

Our current OMM implementation stores obstacles indefinitely. For real-world deployment with temporary obstacles (e.g., stalled vehicles that are towed away), a memory decay mechanism is needed. Potential approaches:
\begin{itemize}
    \item \textbf{Timestamp-based expiry}: Remove obstacles from $O_i(t)$ if not re-observed within time window $\Delta t$
    \item \textbf{Confirmation-based persistence}: Obstacles remain only if periodically re-broadcast by encountering vehicles
    \item \textbf{Probabilistic softening}: Reduce but don't eliminate the cost of routing through old obstacles, allowing reconsideration over time
\end{itemize}

These extensions would allow OMM to distinguish persistent blockages (requiring indefinite avoidance) from temporary delays (permitting eventual re-exploration).

\subsubsection{Security and Trust}

A critical concern for real-world OMM deployment is security against malicious agents. A bad actor could broadcast false obstacle information, causing vehicles to avoid perfectly viable routes and inducing artificial congestion. Mitigations include:
\begin{itemize}
    \item \textbf{Cryptographic authentication}: V2V messages digitally signed with vehicle credentials, enabling verification of sender identity
    \item \textbf{Plausibility checks}: Cross-reference obstacle reports with onboard sensor data; reject reports inconsistent with direct observations
    \item \textbf{Consensus-based validation}: Require multiple independent agents to report an obstacle before adding to $O_i(t)$
    \item \textbf{Reputation systems}: Track message reliability over time; downweight information from agents with history of inaccurate reports
\end{itemize}

Existing V2X security protocols \cite{muslam2024enhancing} provide foundation for these mechanisms, though integration with OMM-specific logic requires further research.

\subsection{Limitations and Future Work}

\subsubsection{Graph Abstraction}

Our graph-based simulation abstracts away vehicle dynamics, acceleration, lane-changing, and intersection protocols. While this enables clear isolation of routing decision effects, it limits direct applicability to real-world scenarios. Future work should validate OMM in high-fidelity simulators (CARLA, SUMO) incorporating detailed vehicle physics and sensor uncertainty. We expect the core benefit—preventing routing loops—to persist, though absolute performance numbers will differ.

\subsubsection{Obstacle Model}

We model obstacles as complete node blockages persisting for fixed durations. Real-world obstacles (accidents, construction, congestion) vary in severity (partial vs. complete blockage), persistence (temporary vs. long-term), and predictability (scheduled construction vs. sudden incidents). Extending OMM to handle probabilistic obstacle models and partial blockages would enhance realism. For instance, obstacles could be represented as $(node, severity, confidence)$ tuples, with routing costs adjusted proportionally rather than binary inclusion/exclusion.

\subsubsection{Heterogeneous Agent Capabilities}

Our experiments assume all vehicles are OMM-capable and communicating. In transitional deployment phases, fleets will be mixed: some vehicles with full OMM, others with partial or no coordination capabilities. Investigating performance degradation as OMM adoption rate varies (e.g., 25

\subsubsection{Scalability to Large Networks}

Our 86-node graph, while sufficient for observing routing loops and OMM effectiveness, is small compared to real metropolitan road networks (thousands to tens of thousands of nodes). Future work should evaluate OMM's scalability to city-sized networks, potentially incorporating hierarchical routing (planning at multiple resolution levels) to manage computational complexity.

\subsubsection{Integration with Learning-Based Approaches}

Our OMM protocol is rule-based, with hand-crafted logic for obstacle memory and path planning. An exciting direction is integrating OMM with MARL. Agents could learn when to trust obstacle information, how long to retain memory, and how to balance exploration (attempting recently-blocked routes to check if cleared) with exploitation (avoiding known obstacles). Preliminary experiments suggest that providing MARL agents with OMM-like memory structures significantly accelerates learning and improves final policy performance, but comprehensive study is future work.

\section{Conclusion}

This research makes three principal contributions to the understanding of multi-agent coordination for autonomous vehicles:

\textbf{First}, we identify and characterize the \textit{routing loop problem}—a pathological failure mode where memory-less reactive rerouting causes vehicles to become trapped in cycles of inefficient path recalculation. Our comprehensive experiments demonstrate that this failure is not an edge case but a systematic phenomenon, causing 190-682\% performance degradation across diverse scenarios. This finding challenges the assumption that more information and more adaptive decision-making necessarily improve multi-agent system performance, highlighting that naive implementations can produce worse outcomes than simpler non-adaptive approaches.

\textbf{Second}, we introduce \textit{Object Memory Management (OMM)}—an elegant, lightweight mechanism for preventing routing loops through persistent shared memory of encountered obstacles. OMM's simplicity (requiring only 12-16 bytes per obstacle message and <200 bytes memory per vehicle) belies its effectiveness: it achieves 70-90\% reductions in delay compared to memory-less systems, bringing performance within 68\% of obstacle-free optimal conditions. By transforming vehicles from reactive agents into proactive, learning agents that leverage collective experience, OMM represents a fundamental advancement in decentralized coordination capability.

\textbf{Third}, through 72 systematically designed experimental configurations spanning diverse vehicle densities, obstacle frequencies, and movement patterns, we provide rigorous empirical evidence that \textit{persistent shared memory is not merely beneficial but essential for robust multi-agent coordination in dynamic environments}. This finding has implications extending beyond autonomous vehicles to general multi-agent AI systems, informing the design of any decentralized coordination mechanism operating under uncertainty.

The transition from individual autonomy to collective intelligence in AV systems requires more than sophisticated sensors and communication protocols—it demands careful algorithm design that accounts for the challenges of decentralized decision-making in non-stationary environments. Our work demonstrates that simple, theoretically-grounded mechanisms like OMM can unlock the full potential of vehicle cooperation, paving the way for resilient, efficient next-generation transportation networks.

As autonomous vehicle technology approaches widespread deployment, insights from this research will inform the practical design of V2V coordination protocols, ensuring that the promise of cooperative driving is realized without the pitfalls of naive implementations. Future research integrating OMM with learning-based approaches, validating performance in high-fidelity simulators, and extending to mixed human-AV traffic will further bridge the gap between theoretical multi-agent coordination and real-world autonomous mobility systems.

\section*{Acknowledgments}

The author expresses deep gratitude to Dr. Daniel Palmer for invaluable mentorship, guidance, and unwavering support throughout this research project. Acknowledgment is also given to the family and friends whose encouragement made this work possible. Finally, thanks to the generative AI tools that assisted with literature summarization and manuscript refinement.

\appendix

\noindent
  \textbf{Table I: Complete Performance Dataset: All 72 Experimental Configurations}
\label{tab:complete_results}
\vspace{0.2cm}

\noindent\resizebox{\textwidth}{!}{%
\begin{tabular}{|l|l|l|ccc|ccc|ccc|ccc|ccc|ccc|}
\hline
\multirow{2}{*}{\textbf{Cars}} & \multirow{2}{*}{\textbf{Obs}} & \multirow{2}{*}{\textbf{Pattern}} & \multicolumn{3}{c|}{\textbf{No Obs}} & \multicolumn{3}{c|}{\textbf{Obs}} & \multicolumn{3}{c|}{\textbf{Comm}} & \multicolumn{3}{c|}{\textbf{OMM}} & \multicolumn{3}{c|}{\textbf{Self}} & \multicolumn{3}{c|}{\textbf{Self+OMM}} \\
\cline{4-21}
& & & T & W & R & T & W & R & T & W & R & T & W & R & T & W & R & T & W & R \\
\hline
15 & 6 & LR & 23.1 & 0.0 & 0 & 39.1 & 16.0 & 0 & 35.0 & 9.4 & 2 & 27.6 & 3.3 & 1 & 51.6 & 13.9 & 4 & 28.1 & 2.7 & 2 \\
15 & 6 & Rand & 14.8 & 0.0 & 0 & 21.5 & 6.7 & 0 & 20.6 & 4.0 & 1 & 19.1 & 2.7 & 1 & 75.5 & 39.6 & 6 & 21.0 & 2.1 & 1 \\
15 & 20 & LR & 23.5 & 0.0 & 0 & 53.4 & 30.0 & 0 & 57.0 & 28.7 & 3 & 39.8 & 14.7 & 2 & 158.2 & 96.3 & 14 & 40.5 & 10.7 & 2 \\
15 & 20 & Rand & 17.5 & 0.0 & 0 & 43.5 & 26.1 & 0 & 51.2 & 29.4 & 4 & 50.5 & 21.4 & 4 & 207.1 & 139.0 & 20 & 51.1 & 18.7 & 3 \\
\hline
35 & 6 & LR & 21.9 & 0.0 & 0 & 34.2 & 12.3 & 0 & 26.8 & 3.7 & 2 & 24.5 & 1.4 & 1 & 43.4 & 14.7 & 4 & 25.0 & 1.2 & 1 \\
35 & 6 & Rand & 16.0 & 0.0 & 0 & 26.3 & 10.3 & 0 & 20.2 & 3.4 & 1 & 18.6 & 1.4 & 1 & 67.0 & 36.2 & 6 & 19.2 & 1.1 & 1 \\
35 & 20 & LR & 22.2 & 0.0 & 0 & 43.0 & 20.9 & 0 & 46.3 & 21.5 & 5 & 40.5 & 15.8 & 2 & 173.4 & 109.8 & 16 & 40.5 & 12.4 & 2 \\
35 & 20 & Rand & 18.8 & 0.0 & 0 & 35.9 & 17.2 & 0 & 39.3 & 18.3 & 3 & 31.1 & 10.0 & 1 & 103.5 & 58.2 & 9 & 29.9 & 7.1 & 1 \\
\hline
55 & 6 & LR & 23.3 & 0.0 & 0 & 35.5 & 12.4 & 0 & 29.5 & 5.1 & 2 & 24.6 & 0.7 & 1 & 98.5 & 56.3 & 9 & 24.7 & 0.6 & 1 \\
55 & 6 & Rand & 14.4 & 0.0 & 0 & 23.4 & 9.1 & 0 & 18.9 & 3.8 & 1 & 17.0 & 1.1 & 1 & 74.8 & 40.7 & 6 & 17.9 & 0.9 & 1 \\
55 & 20 & LR & 23.8 & 0.0 & 0 & 55.4 & 31.7 & 0 & 53.9 & 25.7 & 6 & 45.1 & 17.3 & 3 & 186.0 & 118.9 & 18 & 45.2 & 13.7 & 3 \\
55 & 20 & Rand & 17.1 & 0.0 & 0 & 38.9 & 21.7 & 0 & 38.1 & 17.1 & 5 & 35.6 & 13.5 & 2 & 148.3 & 93.1 & 15 & 35.8 & 10.8 & 2 \\
\hline
\end{tabular}%
}

\vspace{0.2cm}
\noindent\textit{Legend: T = Travel Time (s), W = Wait Time (s), R = Recalculations. LR = Left-to-Right, Rand = Random.}


\begin{thebibliography}{00}
\bibitem{stern2018dissipation} R. E. Stern et al., ``Dissipation of stop-and-go waves via control of autonomous vehicles: Field experiments,'' \textit{Transportation Research Part C: Emerging Technologies}, vol. 89, pp. 205--221, 2018.
\bibitem{kaufeld2024investigating} M. Kaufeld, R. Trauth, and J. Betz, ``Investigating driving interactions: A robust multi-agent simulation framework for autonomous vehicles,'' in \textit{2024 IEEE Intelligent Vehicles Symposium (IV)}, 2024, pp. 803--810.
\bibitem{cui2021scalable} J. Cui et al., ``Scalable multiagent driving policies for reducing traffic congestion,'' \textit{arXiv preprint arXiv:2103.00058}, 2021.
\bibitem{dinneweth2022multi} J. Dinneweth et al., ``Multi-agent reinforcement learning for autonomous vehicles: A survey,'' \textit{Autonomous Intelligent Systems}, vol. 2, no. 1, p. 27, 2022.
\bibitem{yoshizawa2023survey} T. Yoshizawa et al., ``A survey of security and privacy issues in v2x communication systems,'' \textit{ACM Computing Surveys}, vol. 55, no. 9, pp. 1--36, 2023.
\bibitem{tie2024v2x} H. L. Tie, ``V2X vehicle-to-everything communication – the future of autonomous connectivity,'' \textit{Keysight Blogs}, Oct. 2024. [Online]. Available: https://www.keysight.com/blogs/en/inds/auto/2024/10/03/v2x-post
\bibitem{billington2018cv2x} J. Billington, ``C-V2X VS DSRC: Which technology is better for autonomous vehicles?'' \textit{Autonomous Vehicle International}, Oct. 2018.
\bibitem{muslam2024enhancing} M. M. A. Muslam, ``Enhancing security in vehicle-to-vehicle communication: A comprehensive review of protocols and techniques,'' \textit{Vehicles}, vol. 6, no. 1, pp. 450--467, 2024.
\bibitem{zhang2023learning} Y. Zhang et al., ``Learning a robust multiagent driving policy for traffic congestion reduction,'' \textit{Neural Computing and Applications}, pp. 1--14, 2023.
\bibitem{vitale2024cooperative} F. Vitale and C. Roncoli, ``Cooperative Traffic Dispersion through Rerouting of Connected and Automated Vehicles in Urban Networks,'' \textit{Transportation Research Record}, p. 03611981241248648, 2024.
\bibitem{gan2024multi} Q. Gan et al., ``Multi-Vehicle Cooperative Decision-Making in Merging Area Based on Deep Multi-Agent Reinforcement Learning,'' \textit{Sustainability}, vol. 16, no. 22, 2024.
\bibitem{zhou2022multi} W. Zhou et al., ``Multi-agent reinforcement learning for cooperative lane changing of connected and autonomous vehicles in mixed traffic,'' \textit{Autonomous Intelligent Systems}, vol. 2, no. 1, p. 5, 2022.
\bibitem{dijkstra2022note} E. W. Dijkstra, ``A note on two problems in connexion with graphs,'' in \textit{Edsger Wybe Dijkstra: his life, work, and legacy}, 2022, pp. 287--290.
\bibitem{hart1968formal} P. E. Hart, N. J. Nilsson, and B. Raphael, ``A formal basis for the heuristic determination of minimum cost paths,'' \textit{IEEE Transactions on Systems Science and Cybernetics}, vol. 4, no. 2, pp. 100--107, 1968.
\bibitem{chen2022rrt} R. Chen, J. Hu, and W. Xu, ``An RRT-Dijkstra-based path planning strategy for autonomous vehicles,'' \textit{Applied Sciences}, vol. 12, no. 23, p. 11982, 2022.
\bibitem{djavadian2020multiobjective} S. Djavadian et al., ``Multi-objective eco-routing for dynamic control of connected \& automated vehicles,'' \textit{Transportation Research Part D: Transport and Environment}, vol. 87, p. 102513, 2020.
\bibitem{dosovitskiy2017carla} A. Dosovitskiy et al., ``CARLA: An Open Urban Driving Simulator,'' in \textit{Proceedings of the 1st Annual Conference on Robot Learning}, 2017, pp. 1--16.
\bibitem{lopez2018microscopic} P. A. Lopez et al., ``Microscopic Traffic Simulation using SUMO,'' in \textit{The 21st IEEE International Conference on Intelligent Transportation Systems}, 2018.
\end{thebibliography}
\end{document}